%% file: main.tex
\documentclass[conference]{IEEEtran}
\IEEEoverridecommandlockouts
\usepackage{cite}
\usepackage{amsmath,amssymb,amsfonts}
\usepackage{algorithmic}
\usepackage{graphicx}
\usepackage{textcomp}
\usepackage{xcolor}

\input{src/commands}
\providecommand{\Description}[1]{}

\def\BibTeX{{\rm B\kern-.05em{\sc i\kern-.025em b}\kern-.08em
    T\kern-.1667em\lower.7ex\hbox{E}\kern-.125emX}}
\begin{document}

% \title{Conference Paper Title*\\
% {\footnotesize \textsuperscript{*}Note: Sub-titles are not captured for https://ieeexplore.ieee.org  and

\title{MoL-SQL: A Controlled Benchmark for Mixture-of-Language Text-to-SQL}

% \and
% \and
% \and
% \and
% \and

\author{
\IEEEauthorblockN{Tianqi Luo, Boyan Li, Yuyu Luo}
\IEEEauthorblockA{
The Hong Kong University of Science and Technology (Guangzhou)\\
Guangzhou, China\\
tluo553@connect.hkust-gz.edu.cn,
bli303@connect.hkust-gz.edu.cn,
yuyuluo@hkust-gz.edu.cn
}
}

\maketitle

\begin{abstract}
Text-to-SQL systems are usually built and evaluated under a monolingual
assumption, yet in international applications users often query databases
stored in a different language. Existing multilingual benchmarks mostly
translate only the question without aligning schema and value entities, so
they can neither locate the source of an accuracy drop nor rule out
translation errors.
Hence, we introduce \molsql, a benchmark for mixture-of-language Text-to-SQL
that switches the languages of the question, schema, and stored values
independently while keeping each task fixed, and grades every cross-lingual
entity mapping into four difficulty levels.
We migrate five Text-to-SQL benchmarks of different domains and source
languages through a unified alignment pipeline that builds a shared
cross-lingual entity map, migrates each database with its constraints,
rewrites the gold SQL under execution-equivalence checks, and regenerates the
question from the verified SQL, keeping all language versions of a task
aligned in entities, intent, and execution result.
\molsql provides two splits: \molfull ($16{,}892$ samples) for overall
evaluation and \molcube ($3{,}840$ samples) covering every language
combination to isolate question, schema, and value effects.
Experiments with eight LLMs and four families of Text-to-SQL methods show that
schema and value languages have distinct effects, with stored values costing
up to 23 points, and that these effects mainly stem from cross-lingual entity
mapping. Neither question translation nor
additional reasoning narrows the value gap, whereas access to target-database
values substantially reduces it, indicating that mixture-of-language
Text-to-SQL benefits most from language-aware context grounding.
\end{abstract}


\input{src/intro}
\input{src/overview}
\input{src/design-detail}
\input{src/data}
\input{src/exp}
\input{src/related}
\input{src/conclusion}

% 如果使用了需要声明的生成式 AI 内容，在这里添加
% \section*{Acknowledgment}
% ...

\section*{Acknowledgment}

Qwen3-Plus is part of the benchmark construction in Section~III: it
generates the entity map (Section~III-B), realizes the target-language
questions from verified SQL (Section~III-C), and pre-labels the difficulty
level of each entity mapping (Section~III-D1). Its outputs are accepted only
after the deterministic checks and execution check of Section~III, the
difficulty labels after human review, and a sample of the questions is
evaluated by bilingual annotators (Section~IV-C).
The authors used OpenAI Codex to assist with the collaborative development
and modification of code underlying Sections~III--V and their figures.
OpenAI ChatGPT was used for language polishing throughout the manuscript
after the authors prepared the initial drafts. All AI-assisted outputs were
critically reviewed and repeatedly revised by the authors, who take full
responsibility for the content and results.

\bibliographystyle{IEEEtran}
\bibliography{sample}

% \vspace{12pt}
% \color{red}
% IEEE conference templates contain guidance text for composing and formatting conference papers. Please ensure that all template text is removed from your conference paper prior to submission to the conference. Failure to remove the template text from your paper may result in your paper not being published.

\end{document}

%% file: src/commands.tex
\newcommand{\eat}[1]{}

\usepackage{latexsym}
\usepackage{amsfonts}
\usepackage{amsmath}
\usepackage{xcolor}
\usepackage{colortbl}
\usepackage{epsfig}
\usepackage{xspace}
\usepackage{graphicx}
\usepackage{subfigure}
\usepackage{paralist}
\usepackage{enumerate}

\usepackage{amssymb}
\usepackage{enumitem}
\usepackage[color,matrix,arrow,all]{xy}
\usepackage{comment}
\usepackage{booktabs}
\usepackage{balance}
\usepackage{stmaryrd}
\usepackage{pifont}
\usepackage{hhline}
\usepackage{listings}
\usepackage{array}
\usepackage{float}
\usepackage[flushleft]{threeparttable}

\usepackage{mathrsfs}
\usepackage{makecell}
\usepackage{xparse}
\usepackage{epsfig}
\usepackage{multirow}
\usepackage[hyphens]{url}

\usepackage{multirow}
\usepackage{graphicx}
\usepackage{subfigure}
\usepackage{svg}

\usepackage{listings}
\usepackage{framed}
\usepackage{xcolor}
\usepackage{color}
\colorlet{shadecolor}{gray!20}
\usepackage{makecell}

\usepackage{algorithm}
\usepackage{algorithmic}

\definecolor{shadecolor}{RGB}{220,220,220}

\definecolor{lightgreen}{HTML}{D5E8D4}
\definecolor{lightyellow}{HTML}{fbe7cf}
\definecolor{inputcolor}{RGB}{255,139,35}
\definecolor{outputcolor}{RGB}{120,212,252}
\definecolor{embedcolor}{RGB}{254,127,156}
\definecolor{maskcolor}{RGB}{122,128,255}
\definecolor{ecolor}{RGB}{58,149,54}

\definecolor{highcolor}{RGB}{255,153,153}
\definecolor{midcolor}{RGB}{255,204,204}
\definecolor{lowcolor}{RGB}{204,229,255}

\definecolor{srcSpider}{HTML}{4E79A7}
\definecolor{srcKaggle}{HTML}{59A14F}
\definecolor{srcBIRD}{HTML}{E15759}
\definecolor{srcBULL}{HTML}{F28E2B}
\definecolor{srcEHRSQL}{HTML}{B07AA1}
\definecolor{molFullColor}{HTML}{4E79A7}
\definecolor{molCubeColor}{HTML}{E15759}
\definecolor{stageblue}{HTML}{EAF1FA}
\definecolor{stagegray}{HTML}{F2F2F2}

\usepackage{tikz}
\usetikzlibrary{shapes,decorations.pathmorphing}
\usetikzlibrary{calc}
\usetikzlibrary{positioning,arrows.meta,fit,backgrounds}

\usepackage{algorithm} 
\usepackage{algorithmic}  
\usepackage[algo2e]{algorithm2e} 

\usepackage[export]{adjustbox}% http://ctan.org/pkg/adjustbox

\definecolor{red}{RGB}{184,84,80}

\definecolor{green}{RGB}{0,128,0}

\definecolor{yellow}{RGB}{255,200,18}

\definecolor{blue}{RGB}{33,89,202}

\makeatletter
\expandafter\let\expandafter\ltxselectfont\csname selectfont \endcsname
\usepackage{CJKutf8}
\expandafter\let\csname selectfont \endcsname\ltxselectfont
\AddToHook{selectfont}{%
  \expandafter\ifx\csname CJK@\curr@fontshape\endcsname\relax\else
    \CJK@bold@false
    \csname CJK@\curr@fontshape\endcsname
  \fi}
\makeatother
\newcommand{\zh}[1]{\begin{CJK}{UTF8}{gkai}#1\end{CJK}}

\newcommand{\bi}{\begin{itemize}}
\newcommand{\ei}{\end{itemize}}

\newcommand{\be}{\begin{enumerate}}
\newcommand{\ee}{\end{enumerate}}
\newcommand{\beqn}{\begin{eqnarray*}}
\newcommand{\eeqn}{\end{eqnarray*}}

\newcommand{\molsql}{{MoL-SQL}\xspace}
\newcommand{\molmigrate}{{MoL-Migration}\xspace}
\newcommand{\molfull}{{MoL-Full}\xspace}
\newcommand{\molcube}{{MoL-Cube}\xspace}

\usepackage{pifont} % For dingbats (check marks, X marks, etc.)
\newcommand{\cmark}{\ding{51}} % Check mark
\makeatletter
    \newcommand\figcaption{\def\@captype{figure}\caption}
    \newcommand\tabcaption{\def\@captype{table}\caption}
\makeatother

\tikzstyle{mybox} = [draw=black, fill=black!5, thick,
   rectangle, rounded corners, inner sep=0pt, inner ysep=6pt]
\tikzstyle{fancytitle} =[fill=black, text=white]

\NewDocumentCommand{\nan}{ mO{} }{\textcolor{blue}{\textsuperscript{\textit{Nan}}\textsf{\textbf{\small[#1]}}}}

\NewDocumentCommand{\yuyu}{ mO{} }{\textcolor{green}{\textsuperscript{\textit{Yuyu}}\textsf{\textbf{\small[#1]}}}}

\NewDocumentCommand{\leixian}{ mO{} }{\textcolor{blue}{\textsuperscript{\textit{Leixian}}\textsf{\textbf{\small[#1]}}}}

\NewDocumentCommand{\fanj}{ mO{} }{\textcolor{red}{\textsuperscript{\textit{Ju}}\textsf{\textbf{\small[#1]}}}}

\NewDocumentCommand{\tianqi}{ mO{} }{\textcolor{violet}{\textsuperscript{\textit{Tianqi}}\textsf{\textbf{\small[#1]}}}}

\NewDocumentCommand{\wei}{ mO{} }{\textcolor{blue}{\textsuperscript{\textit{Wei}}\textsf{\textbf{\small[#1]}}}}

\usepackage[most]{tcolorbox}

\tcbuselibrary{breakable}

\newtcolorbox{findingbox}{
 enhanced,
 colback=gray!10,
 colframe=black!60,
 boxrule=0.45pt,
 arc=1.5pt,
 left=4pt,
 right=4pt,
 top=3pt,
 bottom=3pt,
 before skip=6pt,
 after skip=6pt,
 fontupper=\small
}

\newtcolorbox{examplebox}[1]{
 enhanced,
 breakable, % 允许盒子跨页
 colback=gray!10,
 colframe=gray!70,
 width=\columnwidth,
 boxrule=0.4pt,
 left=0mm,
 right=0mm,
 top=0mm,
 bottom=0mm,
 before skip balanced=0pt,
 after skip balanced=0pt,
 arc=5pt,
 fonttitle=\bfseries,
 title=#1,
 coltitle=white,  % 使用coltitle替代color
 oversize
}

\usepackage{listings}
\usepackage{xcolor}  % for coloring the code

\lstdefinelanguage{JSON}{
    string=[s]{"}{"},
    stringstyle=\color{red},
    comment=[l]{:},
    commentstyle=\color{blue},
    morecomment=[l]{,},
    morecomment=[l]{[},
    morecomment=[l]{]},
    morecomment=[l]{\{},
    morecomment=[l]{\}},
    keywordstyle=\color{blue},
    keywords={true, false, null}
}

\lstdefinestyle{jsonstyle}{
  language=JSON,
  basicstyle=\rmfamily\small,  % Set the font to typewriter
  keywordstyle=\color{blue},   % Highlight keywords in blue
  stringstyle=\color{red},     % Highlight strings in red
  showstringspaces=false,      % Don't show spaces in strings
  commentstyle=\color{green},  % Highlight comments in green
  breaklines=false,             % Automatically break lines
  numbers=none                 % No line numbers
}

\usepackage{soul}

\definecolor{verylightgray}{rgb}{0.95,0.95,0.95} % Much lighter gray

%% file: src/intro.tex
\section{Introduction}
\label{sec:intro}

Natural-language interfaces let users query and analyze data without writing
code~\cite{ozcan2020NLIDTutorial}. Text-to-SQL, which translates a
natural-language question into an executable SQL query, is the central
instance of this vision, and data agents built on large language models
(LLMs) are making it the default entry point to enterprise
data~\cite{zhu2025SurveyDataAgents,li2026DataAgentsPanel}.
Progress on the task has been driven by increasingly demanding benchmarks:
Spider introduced cross-domain, multi-table evaluation, and BIRD added large,
value-rich databases that require grounding a question in the stored
content~\cite{yu2019SpiderLargeScaleHumanLabeled,li2023CanLLMAlready,
zheng-etal-2024-archer}.
These benchmarks, and the systems developed on
them~\cite{talaei2024chess,li2025omnisql}, share \emph{a monolingual
assumption: the question, the schema identifiers, and the stored values are
written in one language, usually English}.

This assumption does not hold in many deployed databases.
Stored values are frequently multilingual in international business systems
and social-media platforms~\cite{li2024DawnNaturalLanguage,
luo2025NaturalLanguageSQL,alibaba2024PolarDBNL2SQL,
kumar2025EnterpriseGradeNL2SQL,leiSpider20Evaluating2025},
and schemas themselves may be localized, as in government and smart-city
databases~\cite{deng2026TACOBenchmarkOpenDomain,zhang-etal-2023-css}.
Users then ask in one language about data stored in another.
We refer to this setting as \textbf{Mixture-of-Language Text-to-SQL}: the
question, the schema, and the stored values of the same instance may each be
in a different language.
This task is more than question translation.
First, languages differ in how they express the same intent: conditions,
temporal expressions, and comparisons are phrased differently, and
constraints are often left implicit, so a system must recover the precise
intent behind language-specific phrasing.
Second, a system must ground entities across languages. It has to map
question concepts to schema identifiers in another language, often without
any lexical overlap, and produce each literal in the exact form stored in the
database, even when the question mentions it in another language, in an
abbreviated form, or by a domain-specific term.
The language of the question, the schema, and the stored values can each
affect accuracy in its own way.

\input{figures/language_coverage}
\input{figures/icde_translation_examples}

\textbf{Existing Text-to-SQL benchmarks do not isolate these effects.}
Figure~\ref{fig:language-coverage} characterizes each benchmark by the
language of its three components: the question $Q$, the stored values
$V$ of the database, and the schema identifiers $S$, with subscripts en and zh denoting English and Chinese.
Prior benchmarks occupy only selected cells, each on its own workload.
Spider and BIRD are monolingual. CSpider and
MultiSpider~\cite{min2019PilotStudyChineseCspider,
dou2023MultiSpiderBenchmarkingMultilingual} pair non-English questions with
English databases without varying the schema or value language,
DuSQL~\cite{wang2020DuSQLLargeScalePragmatic} is monolingual in Chinese,
TACO~\cite{deng2026TACOBenchmarkOpenDomain} evaluates English and Chinese
workloads separately rather than as aligned realizations of the same
instances, and FinSQL~\cite{zhang2024FinSQLModelAgnosticLLMsbased} starts
from a Chinese database and translates only the questions into English.
Because the covered cells come from disjoint workloads, an accuracy
difference between them mixes the effect of language with differences in
domain, SQL complexity, and annotation.
\textbf{Moreover, translation-based construction can itself introduce errors into the benchmark.}
Figure~\ref{fig:translation-examples} shows how CSpider translates an entity in the
question and the gold SQL but not in the database, so the literal matches no
stored value and the gold query returns an empty result for nearly one-fifth
of the instances; such failures reflect the construction rather than the model.
MultiSpider keeps the English entity in both the Chinese question and
the database, and thus never requires a cross-lingual value match.

Closing this gap requires a benchmark that not only isolates how the language
of each component and the domain contribute to cross-lingual difficulty, but
also avoids errors introduced by its own construction. Crafting such a benchmark presents
three challenges.
\textbf{(C1) Controlled language variation.} The languages of the question,
the schema, and the stored values must be switched independently, while the
user intent and the gold SQL semantics stay fixed and the answers remain
semantically equivalent across languages.
\textbf{(C2) Consistent, task-preserving migration.} An entity recurs in
keys, denormalized copies, constraints, the gold SQL, and the question. A
single inconsistent translation silently breaks a join or a
filter, as in CSpider, so every occurrence must be translated consistently and every
realization verified by execution.
\textbf{(C3) Diagnosable difficulty.} The difficulty of a cross-lingual
entity mapping arises from different sources, such as linguistic
conventions, domain knowledge, and database-specific naming conventions. The
benchmark must cover these sources across domains and grade each mapping by
its difficulty, so that failures can be traced to the knowledge a mapping
requires.

\textbf{Our approach.} To address these challenges, we introduce \molsql, a
benchmark for Mixture-of-Language Text-to-SQL built on the idea of fixing the
task and varying the language of each component.
\textbf{\molsql} consists of two complementary splits: \textbf{\molfull} for overall
evaluation ($16{,}892$ realizations) and \textbf{\molcube}, an
English--Chinese $2\times2\times2$ design that switches the language of the
question, the schema, and the stored values independently for fine-grained
diagnosis ($3{,}840$ realizations) (addressing C1).
To construct them, we develop a framework, \textbf{\molmigrate}, that
migrates five established Text-to-SQL benchmarks through four modules:
\emph{entity map construction} and \emph{task-preserving realization}
translate every occurrence of an entity identically and verify every
realization by execution (addressing C2), while \emph{source curation} and \emph{split
composition} cover different domains and source languages and grade each
entity mapping by difficulty (addressing C3).

Experiments with eight LLMs and four families of Text-to-SQL methods show
that the languages of the schema and of the stored values have distinct
effects: a Chinese schema is harder regardless of the question language,
whereas the stored values lower execution accuracy by up to $23$ points only
when an English mention must be matched to a Chinese value. Where no mapping
needs bridging, the gaps vanish, so these effects stem from cross-lingual
entity mapping rather than from errors introduced by benchmark construction. Beyond measuring the
gaps, \molsql enables a component-wise analysis of existing methods. Neither
question translation nor additional reasoning narrows the value gap, while
grounding generation in the target-database values substantially reduces it,
indicating that Mixture-of-Language Text-to-SQL benefits most from
language-aware context grounding.

\textbf{Contributions.} Our main contributions are as follows:
\begin{enumerate}
  \item We formalize Mixture-of-Language Text-to-SQL, in which the languages
  of the question, the schema, and the stored values vary independently, and
  grade each cross-lingual entity mapping by the knowledge needed to bridge
  it (Section~\ref{sec:problem}).
  \item We develop \molmigrate, a construction framework that translates each
  database, its constraints, and the gold SQL through one entity map,
  verifies every realization by execution, and regenerates the question from
  the verified SQL (Section~\ref{sec:method}).
  \item We construct and release \molsql at
  \url{https://github.com/MoL-SQL/MoL-SQL}, comprising \molfull for overall
  evaluation and \molcube for isolating question, schema, and value effects
  (Section~\ref{sec:dataset}).
  \item Experiments with eight LLMs and four method families show that the
  schema and value languages have distinct effects, and attribute the
  reduction of each gap to individual method components
  (Section~\ref{sec:experiments}).
\end{enumerate}

%% file: figures/language_coverage.tex
\begin{figure}[t]
  \centering
  \begingroup
  \newcommand{\covyes}{\cellcolor{lightgreen}\textcolor{green}{\large\cmark}}
  \newcommand{\covno}{\phantom{\textcolor{green}{\large\cmark}}}
  \newcommand{\coveragepanel}[7]{%
    \begin{minipage}[t]{0.325\linewidth}
      \centering
      \footnotesize\textbf{#1}\\[-0.2em]
      \adjustbox{max width=\linewidth}{\footnotesize\textbf{#2}}\\[-0.15em]
      \scriptsize #3\\[0.12em]
      \footnotesize
      \setlength{\tabcolsep}{1pt}
      \renewcommand{\arraystretch}{1.05}
      \adjustbox{max width=\linewidth}{%
      \begin{tabular}{|c|c|c|}
        \hline
        & {\scriptsize \(DB(V_{\mathrm{en}})\)} &
          {\scriptsize \(DB(V_{\mathrm{zh}})\)} \\
        \hline
        \(Q_{\mathrm{en}}\) & #4 & #5 \\
        \hline
        \(Q_{\mathrm{zh}}\) & #6 & #7 \\
        \hline
      \end{tabular}}
    \end{minipage}%
  }

  \noindent\resizebox{\linewidth}{!}{%
  \coveragepanel{(a) English-sourced}
    {Spider/BIRD/EHRSQL/etc.}
    {schema: \(S_{\mathrm{en}}\)}
    {\covyes}{\covno}{\covno}{\covno}
  \hspace{0.05em}
  \coveragepanel{(b) English-sourced}
    {CSpider, MultiSpider}
    {schema: \(S_{\mathrm{en}}\)}
    {\covno}{\covno}{\covyes}{\covno}
  \hspace{0.05em}
  \coveragepanel{(c) Chinese-sourced}
    {DuSQL/CSS/etc.}
    {schema: \(S_{\mathrm{zh}}\)}
    {\covno}{\covno}{\covno}{\covyes}%
  }

  \vspace{0.05em}

  \noindent\resizebox{\linewidth}{!}{%
  \coveragepanel{(d) Chinese-sourced}
    {FinSQL}
    {schema: \(S_{\mathrm{en}}\) + zh docs}
    {\covno}{\covyes}{\covno}{\covyes}
  \hspace{0.05em}
  \coveragepanel{(e) Bilingual-sourced}
    {TACO}
    {schema: \(S_{\mathrm{en}}/S_{\mathrm{zh}}\)}
    {\covyes}{\covno}{\covno}{\covyes}
  \hspace{0.05em}
  \coveragepanel{(f) \molsql (Ours)}
    {aligned variants}
    {schema: \(S_{\mathrm{en}}/S_{\mathrm{zh}}\)}
    {\covyes}{\covyes}{\covyes}{\covyes}%
  }
  \endgroup

  \caption{Language coverage of representative Text-to-SQL benchmarks.
  (a--e)~selected $Q$--$DB$ cells on disjoint workloads;
  (f)~\molsql fills all four cells on aligned instances.}
  \Description{Six $2\times2$ coverage panels of question language versus
  database value language. Panels (a--e) show prior benchmarks covering
  only selected cells on disjoint workloads; panel (f) shows MoL-SQL
  filling all four cells on aligned instances.}
  \label{fig:language-coverage}
\end{figure}

%% file: figures/icde_translation_examples.tex
\begin{figure}[t]
  \centering
  \includegraphics[width=0.95\linewidth]{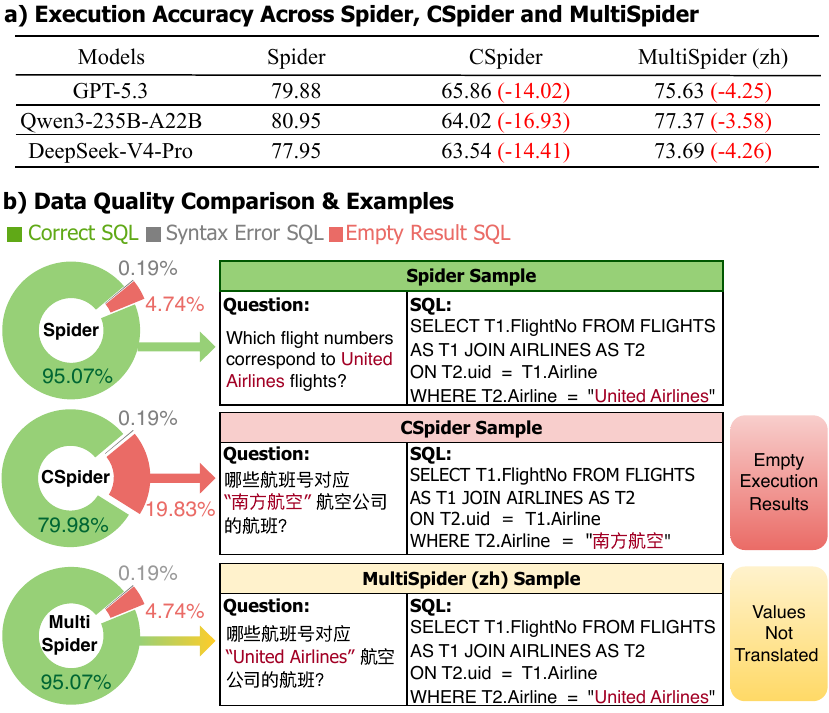}
  \caption{Translation artifacts in CSpider and MultiSpider on
  Spider-dev~\cite{min2019PilotStudyChineseCspider,
  dou2023MultiSpiderBenchmarkingMultilingual}. (a)~Execution accuracy.
  (b)~Rate of gold queries that return an empty result. Empty results in
  CSpider come from entities translated in the question and SQL but not in
  the database.}
  \Description{Two bar charts on Spider-dev. Panel (a) compares execution
  accuracy of CSpider and MultiSpider with the English original; panel (b)
  compares empty-result rates of gold queries.}
  \label{fig:translation-examples}
\end{figure}

%% file: src/overview.tex
\section{Problem Definition}
\label{sec:problem}
\label{sec:overview}

\begin{figure}[t]
    \centering
    \includegraphics[width=\linewidth]{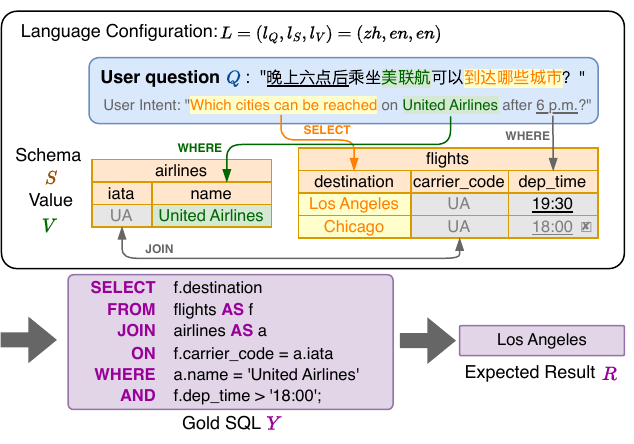}
    \caption{An aligned Mixture-of-Language instance with a Chinese question
    over English schema identifiers and values
    ($L=(\mathrm{zh},\mathrm{en},\mathrm{en})$).}
    \Description{A Mixture-of-Language Text-to-SQL example. A Chinese question
    asks which cities United Airlines reaches after 6 p.m., over English schema
    identifiers and English stored values. The gold SQL joins flights to
    airlines, filters on the stored name and a departure after 18:00, and
    returns Los Angeles.}
    \label{fig:running-example}
  \end{figure}

This section defines the Text-to-SQL setting that \molsql measures and
grades how difficult each cross-lingual entity mapping in an instance is.

\subsection{Problem Formulation}
\label{sec:overview:problem}

Given a natural-language question $Q$ over a database $D=(S,V)$, with schema
$S$ and stored values $V$, a Text-to-SQL system predicts an executable query
$Y$ whose result is $R=\mathrm{Exec}(Y,D)$. The question, the
schema (table and column names), and the stored values are the three
language-bearing components of this input. Conventional benchmarks assume the
three share one language.
\textbf{Mixture-of-Language Text-to-SQL} assigns each component its own language.
A \emph{language configuration} $L=(\ell_Q,\ell_S,\ell_V)$ is that assignment
for the question, the schema, and the translatable stored values.
Choosing a language set for each component gives the configurations as their
product. Section~\ref{sec:problem:config} specializes that product to the
cells the benchmark materializes.

Across these configurations the task stays fixed. A \emph{logical instance}
$i$ fixes the user intent, the database semantics, the meaning of the gold
query, and the answer, independently of language. Its \emph{realization}
under $L$ is $X_i^L=(Q_i^L,D_i^L,Y_i^L)$. Realizations of one instance share a
stable identifier and form paired observations, so an accuracy difference
between them is attributable to the configuration.

Figure~\ref{fig:running-example} shows one realization with
$L=(\mathrm{zh},\mathrm{en},\mathrm{en})$. 
The Chinese question \zh{晚上六点后乘坐美联航可以到达哪些城市？} asks which cities can be reached
on United Airlines after 6~p.m., while the schema identifiers and the stored
values remain in English; the gold query returns Los Angeles.
The question refers to entities of the database in a language different
from the one they are stored in. 
% The mention \zh{城市} (city) denotes the
% column \texttt{flights.destination}, and the airline is reached through the
% join \texttt{flights.carrier\_code = airlines.iata}. The short name
% \zh{美联航} denotes the stored value \texttt{'United Airlines'}, which the
% query must write in its stored English form. The phrase \zh{晚上六点后}
% (after 6~p.m.) becomes the strict predicate \texttt{dep\_time > '18:00'},
% which keeps the 19:30 flight to Los Angeles and excludes the 18:00 flight to
% Chicago.
% Under $L=(\mathrm{zh},\mathrm{en},\mathrm{zh})$, the same instance keeps the
% question and the schema but stores the airline as \zh{美国联合航空}, so the
% same mention must be matched to a different stored form.

\subsection{Cross-lingual Entity Mapping Difficulty}
\label{sec:problem:level}

A language configuration records which components differ in language, but
not how difficult it is to bridge a particular difference. We therefore
consider each \emph{cross-lingual entity mapping}, the correspondence between
a mention in the question and the \emph{entity} it denotes, where an entity
is a table name, a column name, or a stored value referenced by the gold
SQL.\footnote{Key columns that appear only in join conditions are not
counted; the question does not mention them.}
Every instance references at least one table and column, so every instance
contains at least one schema mapping.

The difficulty of a mapping depends on the knowledge required to recover the
stored form from the mention. We define four difficulty levels, summarized in
Table~\ref{tab:mapping-level}. A level is assigned by comparing three
strings: the mention, its \emph{literal gloss} (a word-for-word translation
or transliteration), and the stored form. The conditions are checked from
Level~0 upward, and a mapping receives the lowest level whose condition
holds. Since the mention and the stored form depend on the language
configuration, a mapping is graded separately for each direction
(Section~\ref{sec:method:splits}).

\input{table/icde/tab1_link_taxonomy}

\begin{itemize}
\item \textbf{Level~0 (Invariant).} The mention equals the stored form in
every language. Typical cases are numbers, dates, codes, and identifiers.
\item \textbf{Level~1 (Linguistic).} The literal gloss, or a standard
bilingual dictionary entry, equals the stored form. Typical cases are common
words, names with a standard translation, and schema identifiers.
\item \textbf{Level~2 (Convention).} The stored form follows a convention
outside the database, such as a transliteration, an official name, a domain
term, or a format.
Professional and medical terms often belong to this level: one language may
use a conventional abbreviation where the other spells out the meaning. We
record this asymmetry as \emph{surface opacity}, an attribute independent of
the level.
\item \textbf{Level~3 (Database).} Several renderings are plausible, or the
question does not mention the entity at all, so only the database determines
which form is stored. Typical cases are database-specific codes and names,
rendering variants, pattern matches, and implicit conventions; for example, a
symbol may encode a status, and an abbreviation used only within a table may
stand for a definition fixed by that database.
\end{itemize}

The levels are ordered by the knowledge that a mapping requires. Direct
translation suffices at Level~1, Level~2 requires a stronger model or
additional domain knowledge, and Level~3 requires grounding in the database
that fixes the stored form. Section~\ref{sec:experiments} tests these
correspondences.

%% file: table/icde/tab1_link_taxonomy.tex
% Table I (Sec. II-B), spans both columns. One example per case. Conditions are checked from Level 0 upward; a mapping
% receives the lowest level whose condition holds. Each condition compares the literal gloss with the stored form.
% All examples occur in MoL-SQL questions (Q_zh, S_en/V_en cells; MoL-Cube unless noted): 1980 (BIRD formula_1), 031-19622 (EHRSQL), LifeExpectancy (Spider world_1), English (Spider tvshow), K-9 (BIRD california_schools),
% Ben Jones (Spider tvshow), Star Trek (BIRD superhero, MoL-Full only; 'Star Trek' also in V_zh), HGB and
% Admission = '+' (BIRD thrombosis_prediction; '+' in every cell),
% 1 month and Chairman of the Board (BULL), alive (EHRSQL). See outline/08-link-type-examples.md.
% Thin grey rules separate the levels (\arrayrulecolor needs colortbl; \multirow needs multirow).
\begin{table*}[t]
\centering
\footnotesize
\setlength{\tabcolsep}{2.6pt}
\renewcommand{\arraystretch}{1.12}
\newcommand{\typesep}{\arrayrulecolor{black!25}\midrule[0.3pt]\arrayrulecolor{black}}
\caption{Examples of cross-lingual entity mappings at each difficulty level.}
\label{tab:mapping-level}
\begin{tabular}{@{}l l l l l@{}}
\toprule
\textbf{Level} & \textbf{Typical case} & \textbf{Chinese mention} & \textbf{Literal gloss} &
  \textbf{Stored form} \\
\midrule
\multirow{2}{*}{\textbf{Level 0} Invariant}
  & Numbers, dates & \zh{1980年} & 1980 & \texttt{'1980'} \\
 & Codes, IDs & \zh{患者}\,031-19622 & patient 031-19622 & \texttt{'031-19622'} \\
\typesep
\multirow{2}{*}{\textbf{Level 1} Linguistic}
  & Common words, names & \zh{英语} & English & \texttt{'English'} \\
 & Identifiers & \zh{预期寿命} & life expectancy & \textsf{LifeExpectancy} \\
\typesep
\multirow{4}{*}{\textbf{Level 2} Convention}
  & Transliterated names & \zh{本·琼斯} & Ben Qiongsi & \texttt{'Ben Jones'} \\
 & Official names & \kern-0.5em\zh{《星际迷航》} & lost course among the stars & \texttt{'Star Trek'} \\
 & Domain terms & \zh{血红蛋白} & hemoglobin & \textsf{HGB} \\
 & Format, meaning & \zh{幼儿园到九年级} & kindergarten to ninth grade & \texttt{'K-9'} \\
\typesep
\multirow{4}{*}{\textbf{Level 3} Database}
  & DB-specific codes, names & \zh{住院随访} & inpatient follow-up & \texttt{Admission = '+'} \\
 & Rendering variants & \zh{一个月} & one month & \texttt{'1 month'} \\
 & Pattern match & \zh{董事长} & chairman & \texttt{'\%Chairman of the Board\%'} \\
 & Implicit convention & \zh{生存率} & survival rate & \texttt{status = 'alive'} \\
\bottomrule
\end{tabular}
\end{table*}

%% file: src/design-detail.tex
\section{Benchmark Construction: \molmigrate}
\label{sec:method}

\input{figures/icde_module_overview}

To migrate data sources from different domains, at scale, into aligned
realizations of each logical instance, we design \molmigrate.
Three design goals organize the construction:
\textbf{(1) Entity-consistent migration}: all occurrences of an entity
share one translation, and fixed literals map to themselves.
\textbf{(2) Task-preserving realization}: every realization passes an
execution-equivalence check, and questions are generated only from verified
SQL and databases.
\textbf{(3) Diagnosable coverage}: schema and value language can be
switched separately, and an explicit difficulty standard identifies the
source of each question's difficulty for later system optimization.
As Figure~\ref{fig:overview} shows, \molmigrate has four modules,
introduced in the following subsections.

\subsection{Source Curation}
\label{sec:method:curation}

% \textbf{Challenge and invariant.} A benchmark is only as trustworthy as its
% starting point: the gold SQL must be correct, and the sources must cover every
% difficulty level. After Module~A, every retained instance has executable gold SQL and
% a stable identifier, and every database object has an entity type.

\textbf{Source selection.} We select five sources for three reasons.
They are well-established Text-to-SQL benchmarks.
They cover complementary domains: general, cross-domain databases, such as
Spider~\cite{yu2019SpiderLargeScaleHumanLabeled},
BIRD~\cite{li2023CanLLMAlready}, and
KaggleDBQA~\cite{lee2021KaggleDBQA}, and specialized domains, such as finance
(BULL~\cite{zhang2024FinSQLModelAgnosticLLMsbased}) and clinical care
(EHRSQL~\cite{lee2026EHRSQLPracticalTexttoSQL}).
They also differ in emphasis and in their distribution over the difficulty
levels of Section~\ref{sec:problem:level}, so the sources complement
one another.
Spider is expected to be dominated by Level~1, with cross-domain
compositional queries;
BULL elevates convention-driven (Level~2) financial wording and, as a natively
Chinese source, reverses the migration direction;
EHRSQL elevates database-specific clinical names and codes (Level~3);
BIRD and KaggleDBQA mix the three, BIRD through value-rich databases and
KaggleDBQA through abbreviated columns.
% Section~\ref{sec:dataset:level} checks this expectation against the labeled
% data. 
% We start from the original releases rather than from
% CSpider or MultiSpider, which would inherit the artifacts of
% Figure~\ref{fig:translation-examples}.

\textbf{Instance retention.} An instance is retained if its gold SQL
executes on SQLite, parses into an AST, and keeps a language-invariant answer
after migration. Of 4{,}234 indexed candidates, 11 are dropped due to
text ordering, text aggregation, or value collisions, and 31
receive reviewed gold-SQL repairs such as stable tie-breaks, leaving 4{,}223
instances (Table~\ref{tab:statistics}).

\textbf{Entity typing.} Database objects are typed as \emph{translatable}
(table names, column names, natural-language values) or \emph{fixed literals}
(identifiers, short codes, dates, numbers, URLs), using column types, regular
expressions, and a reviewed whitelist. Fixed literals map to themselves in
every realization. The output is the set of retained logical instances with
typed entities.

\subsection{Entity Map Construction}
\label{sec:method:phase1}

In a database, the same entity can appear in
several places---primary and foreign keys, denormalized copies, undeclared join
values with semantically related concepts. 
For example, if one copy of value \texttt{'United Airlines'} is translated differently from another, a join returns nothing while the SQL still runs. 
Module~B ensures that all occurrences in one entity cluster share one translation, fixed literals map to themselves, and no two distinct entities are mapped to the same translation.
\input{figures/icde_schema_graph}

\textbf{Dependency discovery.} We model each database as a column-level
graph $G=(C,E_{\mathrm{exp}}\cup E_{\mathrm{imp}})$. 
As is shown in Figure~\ref{fig:schema-graph}, a graph is constructed for each database.
Explicit edges
$E_{\mathrm{exp}}$ come from declared foreign keys and equivalences; implicit
edges $E_{\mathrm{imp}}$ link columns whose shared distinct values make up at
least 30\% of the smaller column's, which recovers denormalized
copies without merging columns on one coincidental string.
Figure~\ref{fig:schema-graph} shows both kinds on the running schema.
Connected components of $G$ are \emph{entity clusters}, the unit of
translation. Two cases need care: equal column names do not imply the same
entity, and the same string in unrelated columns may denote different
entities; both are resolved by the table and column names in the schema that every entry carries.

\textbf{Map generation.} An LLM (Qwen3-Plus) translates each cluster in
batches of at most 50 values, with
every identifier and value presented in its table and column context; 
a value's translation is copied to every table and column position in its cluster.
% The map is materialized once, so all downstream steps are
% reproducible and can be rerun independently.

\textbf{Map validation.} Deterministic checks cover coverage, residual
source-language strings, and conflicts. A many-to-one translation would break
\texttt{DISTINCT}, \texttt{GROUP BY}, and uniqueness constraints, so conflicts
are regenerated or, if unresolvable, kept as fixed literals. The output is $M=M_S\cup M_V$, with schema and value entries stored separately.

\subsection{Task-preserving Realization}
\label{sec:method:phase3}
\label{sec:method:phase2}
\label{sec:method:phase4}
\label{sec:method:mixed-sv}

In a realization, the database, the gold SQL, and the question must describe
the same task.
However, string replacement breaks SQL scope, and free translation shifts intent.
Therefore, we need to ensure that the schema identifiers, string literals, aliases and CTE names, and the wording of the question are correctly translated.
Module~C ensures that every realization satisfies
$\mathrm{Exec}(Y',D')\cong M(\mathrm{Exec}(Y,D))$, and that its question matches
the intent of $Y'$.

\textbf{Database migration.} $M$ is applied to schema objects, stored
values, \texttt{CHECK} constraints, and views in lockstep. Because
constraints and views are SQL expressions, they are rewritten by the same
engine as the gold queries (scope-aware SQL rewriting, below), so a
constraint such as
\texttt{CHECK (status IN ('Active','Pending'))} stays valid
under translated values. Crucially, $M_S$ and $M_V$ can be applied
separately: applying neither, $M_S$ only, $M_V$ only, or both yields
$(S_{\mathrm{en}},V_{\mathrm{en}})$, $(S_{\mathrm{zh}},V_{\mathrm{en}})$,
$(S_{\mathrm{en}},V_{\mathrm{zh}})$, and $(S_{\mathrm{zh}},V_{\mathrm{zh}})$.
This is what allows \molcube to switch schema and value language
independently. Each variant is checked for missing entities, residual source
strings, name collisions, and constraint violations.

\input{figures/icde_ast_rewrite}

\textbf{Scope-aware SQL rewriting.} The goal is to rewrite the gold query so
that it executes on the migrated database and returns the translated
counterpart of the original result, while SQL scope stays intact. As
Figure~\ref{fig:ast-rewrite} shows, the rewriter parses the gold query into
an AST, qualifies every column reference against the schema before anything
is renamed, rewrites the qualified AST with $M$, and generates the target
query, which is accepted only if it passes the execution check below. The
rewrite renames only physical tables and columns; aliases, CTE names, and
derived columns are kept, and a string literal is replaced only if $M$ has an
entry for the column it is compared with, so the same string in another
column is left unchanged. Ordering and aggregation over text can change
meaning after translation; the instances whose answers changed for this
reason are the drops reported under instance retention. Queries that fail to
parse fall back to a regular-expression rewriter and, if that fails, to
manual repair.

\textbf{Execution check.} Acceptance is decided by execution, not by
surface form.
Let $M$ act cellwise on query results. A rewritten query $Y'$ on the migrated
database $D'$ is \emph{execution-equivalent} to $Y$ on $D$ if
$\mathrm{Exec}(Y',D')\cong M(\mathrm{Exec}(Y,D))$, where $\cong$ is ordered
equality when the query orders its result and multiset equality otherwise,
duplicates and \texttt{NULL}s are preserved, and errors, timeouts, and arity
mismatches are never equivalent.
Every database variant of every instance must pass this check. A failure
returns to Module~B to correct $M$ or, if no correct map exists, removes the
instance from the funnel. The check is applied to the released database state;
we do not claim semantic equivalence over all states.

\textbf{Question realization.} The question is generated from the verified
$(D',Y')$, not translated freely: the LLM receives the target-language tables,
columns, literals, and comparison operators used by $Y'$, together with the
original question to keep its implicit conditions and style. Checks verify
that referenced entities appear, comparison words are preserved, and the
question uses the mention that $M$ assigns to each value; mismatches trigger
regeneration.
We release the generated questions, since a temperature of zero does not make
generation reproducible across model versions.

\textbf{Composition.} A realization
$X_i^L=(Q_i^{\ell_Q},D_i^{\ell_S,\ell_V},Y_i^{\ell_S,\ell_V})$ joins a
question and an execution-checked database variant by the stable instance
identifier, so every composed cell inherits the execution guarantee.

\subsection{Split Composition}
\label{sec:method:splits}

Module~D grades the difficulty of every cross-lingual entity mapping and
releases the retained instances as two splits, \molfull and \molcube.

\subsubsection{Difficulty Annotation}
Each mapping is graded at the entity level, and these levels are aggregated
into instance-level tags.

\textbf{Entity level.} Grading every mapping of every instance by hand does
not scale, and string matching cannot separate Level~1 from Level~2 and
misreads conventions such as \texttt{'K-9'} as Level~3. We therefore combine
deterministic extraction, LLM judgment, and human revision. First, the
entities of the gold SQL are extracted from its AST (join-only keys
excluded), and their forms in both languages are read from $M$; a mapping
whose mention equals the stored form receives Level~0 without an LLM call.
Second, for every other mapping the LLM receives both questions and the
entity with its table and column context, and a prompt that contains only the
conditions and examples of Table~\ref{tab:mapping-level}; it returns the
mention span (or ``not mentioned''), the literal gloss, the level and typical
case, and a one-line reason, separately for each direction. Third,
identical mention--entity pairs share one label, two independent runs are
compared, and disagreements and all boundary cases (abbreviations,
database-specific prefixes, compound clinical strings) go to human review.
Prompts, code, and labels are released with the benchmark.

\textbf{Instance level.}
\label{sec:method:tags}
The \emph{instance level} is the highest level among an instance's mappings.
An instance with no value mapping, called \emph{schema-only}, takes the level
of its schema mappings. Value mappings are also graded on their own, so value
difficulty can be read apart from the instance level. Reported instance
levels (Section~\ref{sec:dataset:level}) and experimental strata ignore
mappings to \emph{unmentioned} entities, which the question never refers to.
Such mappings remain a Level-3 case in Table~\ref{tab:mapping-level}, but
they belong to one question--entity pair and would otherwise dominate the
maximum over schema mappings. Further tags record schema-only versus
value-bearing, the number of mapped entities, the grounding load (the number
of mappings above Level~0), and SQL
difficulty, which is identical across the cells of an instance and serves as
a covariate.

\subsubsection{Language Effect and Dataset Splits}
\label{sec:problem:config}
To explain how the language of the question, the schema, and the stored
values changes execution accuracy, we define a \emph{language effect} and use
it as a metric in the experiments (Section~\ref{sec:experiments}).
A language effect is the main effect of one component's language on
execution accuracy (EX). For values,
\begin{equation}
\label{eq:effect}
\Delta_V=\tfrac{1}{4}\textstyle\sum_{q,s}\bigl[\mathrm{EX}(q,s,\mathrm{en})-\mathrm{EX}(q,s,\mathrm{zh})\bigr],
\end{equation}
with $\Delta_Q$ and $\Delta_S$ defined analogously; a positive value means
the English realization is easier. A \emph{mismatch effect} is the
corresponding interaction, such as $Q{\times}V$.

The benchmark is therefore released as two splits.
\begin{itemize}
\item \textbf{\molfull.} The coverage split retains every instance in the four
coupled configurations $\{Q_{\mathrm{en}},Q_{\mathrm{zh}}\}\times
\{DB_{\mathrm{en}},DB_{\mathrm{zh}}\}$, where $DB$ switches $S$ and $V$
together, and preserves the source distribution. It supports overall and
per-source evaluation and identifies $\Delta_Q$ and the coupled database
effect $\Delta_{DB}$.
\item \textbf{\molcube.} The factorial split materializes a source-balanced
subset in all eight cells. The subset is enriched in value-bearing instances
and in mappings above Level~1, so that Levels~2 and~3 stay estimable, and
includes schema-only instances as a control
(Section~\ref{sec:dataset:stats}). Because the three languages are crossed,
$\Delta_Q$, $\Delta_S$, and $\Delta_V$ each isolate one component, and the
interactions among them are mismatch effects.
\end{itemize}

%% file: figures/icde_module_overview.tex
\begin{figure*}[t]
  \centering
  \includegraphics[width=\textwidth]{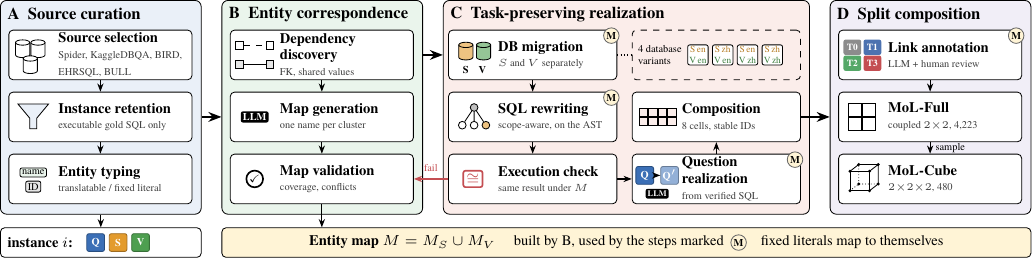}
  \caption{\molmigrate. Each module lists its steps in order. Module B builds
  the entity map $M=M_S\cup M_V$ (bottom bar); steps marked \textsf{M} read it.
  LLM calls occur only in map generation, question realization, and
  difficulty annotation (followed by human review). The only feedback edge sends
  instances that fail the execution check back to Module~B, where the map is
  fixed or the instance is dropped. Modules~B--D serve entity-consistent
  migration, task-preserving realization, and diagnosable coverage.}
  \Description{Four boxes from left to right: A Source curation (source
  selection, instance retention, entity typing), B Entity map construction
  (dependency discovery, map generation, map validation), C Task-preserving
  realization (DB migration, SQL rewriting, execution check, question
  realization, composition), and D Split composition (difficulty annotation,
  MoL-Full, MoL-Cube). A shared entity-map bar runs below B to D.}
  \label{fig:overview}
\end{figure*}

%% file: figures/icde_schema_graph.tex
\begin{figure}[t]
  \centering
  \includegraphics[width=0.75\linewidth]{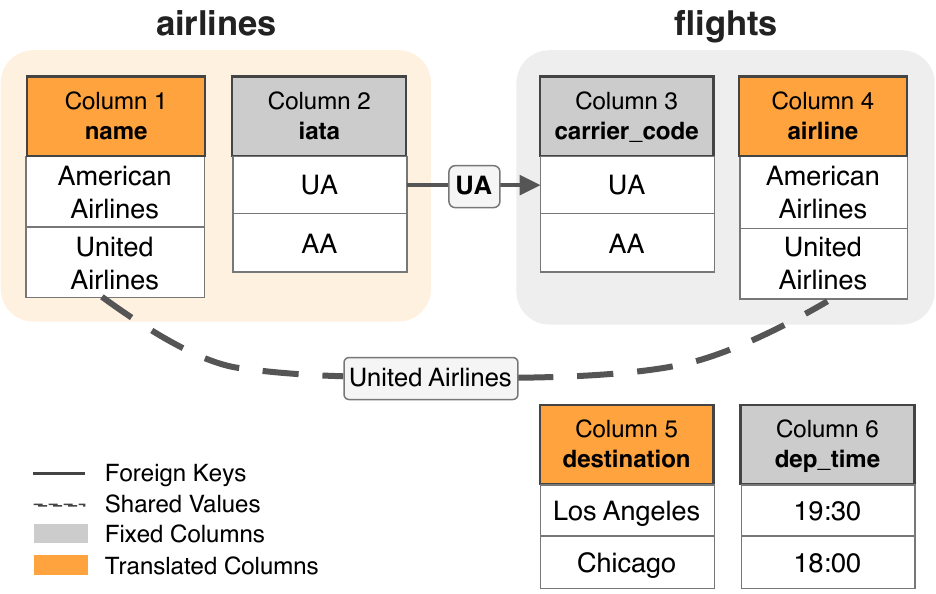}
  \caption{Dependency graph of the running airline schema.}
  \Description{Six column nodes with sample cells. A solid edge links iata and
  carrier\_code, both storing UA. A dashed edge links airlines.name and
  flights.airline, both storing United Airlines.}
  \label{fig:schema-graph}
\end{figure}

%% file: figures/icde_ast_rewrite.tex
\begin{figure*}[t]
  \centering
  \includegraphics[width=\textwidth, keepaspectratio=false]{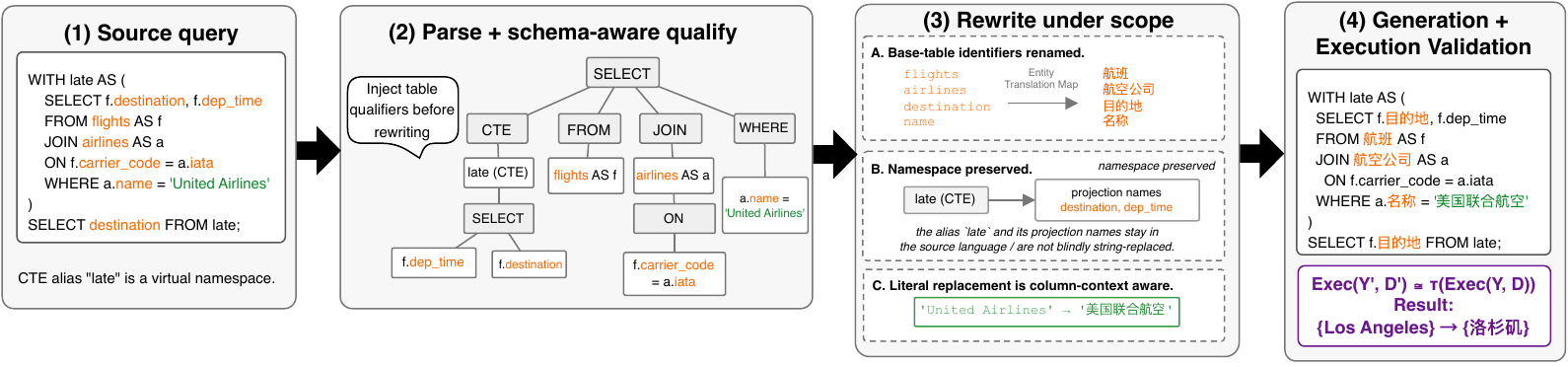}
  \caption{Scope-aware SQL rewriting on the running query.}
  \Description{Four-stage pipeline from an English gold SQL query through a
  qualified AST and scope-aware rewrite to generated Chinese SQL, with an
  execution-equivalence result.}
  \label{fig:ast-rewrite}
\end{figure*}

%% file: src/data.tex
\section{Benchmark Characterization and Quality}
\label{sec:dataset}

\input{table/icde/tab3_statistics_funnel}

Applying \molmigrate (Section~\ref{sec:method}) to the five source workloads
yields the two splits of \molsql. Before using them for measurement, this
section answers three questions: how large each split is and what it covers
(Section~\ref{sec:dataset:stats}), which difficulty levels of cross-lingual
entity mapping it exercises (Section~\ref{sec:dataset:level}), and how
reliable its realizations are
(Section~\ref{sec:dataset:quality}).

\subsection{Scale and Composition}
\label{sec:dataset:stats}
\label{sec:dataset:full}
\label{sec:dataset:cube}

Table~\ref{tab:statistics} reports the construction funnel and the size of
both splits, counting logical instances and realizations separately.
Curation (Section~\ref{sec:method:curation}) leaves 4{,}223 logical
instances over 44 logical databases. Each database is materialized in the
four schema/value language variants of Section~\ref{sec:method:phase3}
(176 materializations), all derived from one entity map.

\textbf{Source splits.} From every source we take a held-out split, never
data released for training or adaptation; the per-source counts are fixed
before migration (Table~\ref{tab:statistics}). Three choices deviate from
the default dev split: for BIRD we use the manually re-verified mini-dev; for
BULL we take the index-aligned English questions that BULL itself provides,
so no BULL question is machine-translated; and for EHRSQL we keep only the
answerable questions, because unanswerable ones carry no gold SQL and
answerability detection is orthogonal to cross-lingual entity mapping.
The only sampling step is the per-source draw of \molcube described below.

\textbf{\molfull} contains all 4{,}223 instances in four coupled
configurations (16{,}892 realizations). It inherits the source mix of the
original workloads, from EHRSQL (35.8\%) and Spider (24.4\%) down to
KaggleDBQA (4.4\%); this mix reflects the source benchmarks, not the
prevalence of language configurations in real workloads.

\textbf{\molcube} selects 96 instances per source (480 instances) from
\molfull and realizes each in all eight cells (3{,}840 realizations), so
that no single workload dominates an effect estimate. Instances are chosen
by a deterministic greedy fill with per-source quotas for value-bearing,
schema-only, and recurring-entity instances, a target difficulty mix,
and a per-database cap; the procedure is released with the benchmark. As a
result, 69\% of \molcube instances contain a value that changes language,
against 21\% in \molfull, while the 117 schema-only instances (24.4\%) form
the control group for the value effect.

\textbf{Entity map.} The last block of Table~\ref{tab:statistics} sizes the
entity map $M$ of Section~\ref{sec:method:phase1}, the only artifact from
which every database variant, rewritten gold query, and value mention is
derived. Its size is therefore the amount of translation that had to stay
consistent across all occurrences and pass the execution check, and we report
it per source so that the migration can be audited. $M_S$ covers every table
and column of the 44 databases (200 and 1{,}600), not only those used by the
gold SQL, so switching the schema language changes the whole schema and the
language of an identifier does not reveal whether a query needs it; BULL needs
no schema entries because its Chinese schema is native. The 11{,}929 value
entries concentrate in BULL, BIRD, and EHRSQL, the sources whose gold SQL
most often compares against stored text, which is where the value axis has
the most material. Finally, 312 items are kept as fixed literals because they fall into the categories of codes or identifiers. They read the same in every realization, showing that migration translates selectively rather than inventing renderings for codes.

% \textbf{Realization properties.} Two naturally occurring phenomena are kept
% rather than filtered out, and both are judged on the actual mention and
% stored form rather than on cell labels (Section~\ref{sec:problem:level}).
% First, some values stay untranslated in a $V_{\mathrm{zh}}$ cell, which
% affects 36/96 EHRSQL instances in \molcube and about
% 42--44\% of BIRD, EHRSQL, and BULL instances in \molfull, so the coupled
% $\Delta_{DB}$ on \molfull is a conservative estimate. Second, Chinese
% questions sometimes keep an English form (code-switching), as bilingual
% users do; 27/96 Spider questions in \molcube do so. Section~\ref{sec:experiments}
% stratifies by code-switching and reports effects with and without it.

\input{figures/icde_linktype_by_source}

\subsection{Distribution of Difficulty Levels}
\label{sec:dataset:level}

Figure~\ref{fig:level-by-source} reports instance levels
(Section~\ref{sec:method:tags}): each instance takes the highest level among
its schema and value mappings, ignoring mappings to entities the question
never mentions; no instance is at Level~0. Twelve \molcube instances and two
\molfull instances are omitted because they contain unaligned literals or
only mappings to unmentioned entities, which leaves 468 and 4{,}221 graded
instances.

The per-source profiles match the rationale for source selection in
Section~\ref{sec:method:curation}. Spider is mostly at Level~1 (90\% on
\molfull, 75\% on \molcube) because its cross-domain vocabulary has
dictionary equivalents. BULL has the largest Level-2 share (68\% on
\molcube, 44\% on \molfull) because financial terms follow established
renderings, and about half of EHRSQL is at Level~3 (49--50\%) because
clinical names and codes are fixed by the database. BIRD and KaggleDBQA mix
the three levels. Each source thus contributes a distinct bridging
situation, which is what lets Section~\ref{sec:experiments} explain
differences between sources by their level mix.

At the split level, \molfull is 48\%/32\%/20\% at Levels~1/2/3. Because
\molcube favors value-bearing and harder instances within each source
(Section~\ref{sec:dataset:stats}), it shifts weight from Level~1 toward
Levels~2 and~3 (44\%/39\%/17\%; 205/184/79 instances). Even the rarest
level, Level~3, keeps 79 instances, although subsets restricted to one
component and one bridging direction can be much smaller
(Section~\ref{sec:exp:difficulty}).

\input{table/icde/tab5_human_audit_template}

\subsection{Quality Assessment}
\label{sec:dataset:quality}

We assess quality in two ways: automatic checks applied to every instance
and a human evaluation of sampled realizations.

\textbf{Automatic checks.} Every retained instance passes the execution
check in every database variant (Section~\ref{sec:method:phase3}), and
candidates without a language-invariant answer are excluded rather than
relaxed. Difficulty annotation also flags
question--map inconsistencies, where the question uses a different rendering
than the migrated value. We corrected these by
regenerating the question or the map entry and keep the flagged instances as
a subset, which Section~\ref{sec:experiments} removes in a sensitivity check.

\textbf{Human evaluation of realizations.} Execution equivalence certifies
the database and the gold SQL, but not whether the regenerated question
still asks for what the SQL computes. We therefore sample 20 \molcube
instances per source (100 instances), balanced over the migration direction
(en$\to$zh for four sources, zh$\to$en for BULL) and the instance level,
and evaluate their target-language realizations. Three bilingual annotators
with SQL experience rate each item independently. We average their ratings
on three criteria scored on a 1--5 scale:
\emph{semantic fidelity}, whether the question preserves the intent,
conditions, and answer semantics of the gold SQL $Y'$;
\emph{naturalness}, whether the question is fluent in its language; and
\emph{entity grounding}, whether schema and value mentions correspond to the
migrated entities. The overall scores are
4.57 for fidelity, 4.60 for naturalness, and 4.27 for entity grounding.
Agreement is above chance on all three criteria (permutation test, $p<0.001$),
with Krippendorff's ordinal $\alpha$ of 0.34/0.26/0.48 and an ICC(2,$k$) of
0.63/0.63/0.73 for the averaged scores. Table~\ref{tab:human-audit} reports the scores by source,
direction, and difficulty level. Together, these results provide multidimensional
evidence for the quality of the generated realizations. 
% These direct
% judgments complement the indirect evidence of Section~\ref{sec:exp:difficulty},
% where switching the value language moves accuracy by at most 3.1 points
% when no value mapping needs bridging.

%% file: table/icde/tab3_statistics_funnel.tex
% Copied from MoL-SQL-ICDE-2027/tables (generated by scripts/build_stats.py).
% Still to recover from build logs: regex-fallback counts.
\begin{table*}[t]
\centering
\small
\setlength{\tabcolsep}{3.6pt}
\caption{Construction funnel, split sizes, and entity-map size per source (Real.: realizations).}
\label{tab:statistics}
\begin{tabular}{@{}l rrrr rr rr r rrrr@{}}
\toprule
 & \multicolumn{4}{c}{\textbf{Funnel}} & \multicolumn{2}{c}{\textbf{\molfull}} & \multicolumn{2}{c}{\textbf{\molcube}} & &
   \multicolumn{4}{c}{\textbf{Entity map}} \\
\cmidrule(lr){2-5}\cmidrule(lr){6-7}\cmidrule(lr){8-9}\cmidrule(l){11-14}
\textbf{Source} & Upstream & Indexed & Dropped & Repaired & Logical & Real. & Logical & Real. & \#DB &
  Tables & Columns & Values & Fixed \\
\midrule
Spider & 1,034 & 1,034 & 4 & 7 & 1,030 & 4,120 & 96 & 768 & 20 & 81 & 441 & 1,251 & 258 \\
KaggleDBQA & 272 & 185 & 1 & 6 & 184 & 736 & 96 & 768 & 8 & 17 & 179 & 699 & 9 \\
BIRD & 500 & 500 & 2 & 8 & 498 & 1,992 & 96 & 768 & 11 & 75 & 796 & 3,393 & 45 \\
EHRSQL & 2,239 & 1,515 & 4 & 2 & 1,511 & 6,044 & 96 & 768 & 2 & 27 & 184 & 1,988 & 0 \\
BULL & 1,000 & 1,000 & 0 & 8 & 1,000 & 4,000 & 96 & 768 & 3 & -- & -- & 4,598 & 0 \\
\midrule
Total & 5,045 & 4,234 & 11 & 31 & 4,223 & 16,892 & 480 & 3,840 & 44 & 200 & 1,600 & 11,929 & 312 \\
\bottomrule
\end{tabular}
\end{table*}

%% file: figures/icde_linktype_by_source.tex
\begin{figure}[t]
  \centering
  \includegraphics[width=\columnwidth]{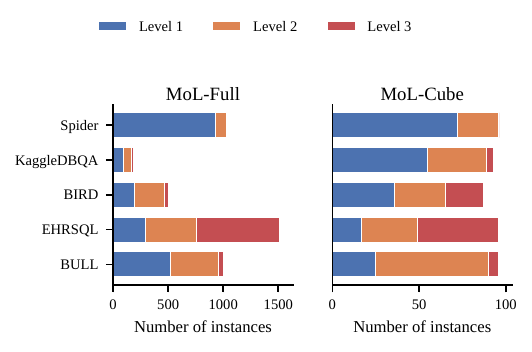}
  \caption{Logical instances by difficulty level and source on \molfull{} (left) and \molcube{} (right).}
  \Description{Two panels of horizontal stacked bars in one column. Left is
  MoL-Full and right is MoL-Cube, with one bar per source. Bar length is the
  number of logical instances, colored by Level 1, Level 2, and Level 3. The
  horizontal scales differ between panels.}
  \label{fig:level-by-source}
\end{figure}

%% file: table/icde/tab5_human_audit_template.tex
% Table (Sec. IV-C): human evaluation of sampled realizations.
% Filled from annotation/human_audit_statistics.json (three reviewers, all 100 items;
% scores: three-reviewer means). Agreement: annotation/human_audit_agreement.json.
\begin{table}[t]
\centering
\small
\setlength{\tabcolsep}{3pt}
\caption{Human evaluation.}
\label{tab:human-audit}
\begin{tabular}{@{}l r ccc@{}}
\toprule
\textbf{Stratum} & \textbf{$n$} & \makecell{\textbf{Fidel-}\\\textbf{ity}} & \makecell{\textbf{Natural-}\\\textbf{ness}} &
\makecell{\textbf{Entity}\\\textbf{ground.}} \\
\midrule
\multicolumn{5}{@{}l}{\emph{By source}} \\
\quad Spider & 20 & 4.45 & 4.65 & 4.02 \\
\quad KaggleDBQA & 20 & 4.37 & 4.67 & 4.55 \\
\quad BIRD & 20 & 4.43 & 4.52 & 4.42 \\
\quad EHRSQL & 20 & 4.82 & 4.63 & 4.38 \\
\quad BULL & 20 & 4.80 & 4.55 & 4.00 \\
\multicolumn{5}{@{}l}{\emph{By direction}} \\
\quad en$\to$zh (4 sources) & 80 & 4.52 & 4.62 & 4.34 \\
\quad zh$\to$en (BULL) & 20 & 4.80 & 4.55 & 4.00 \\
\multicolumn{5}{@{}l}{\emph{By difficulty level}} \\
\quad Level 1 & 48 & 4.56 & 4.58 & 4.47 \\
\quad Level 2 & 27 & 4.59 & 4.59 & 4.04 \\
\quad Level 3 & 25 & 4.57 & 4.65 & 4.15 \\
\midrule
All & 100 & 4.57 & 4.60 & 4.27 \\
\bottomrule
\end{tabular}
\end{table}

%% file: src/exp.tex
\section{Experiments}
\label{sec:experiments}

% We first describe the setup and report the main results of representative
% LLMs and Text-to-SQL methods on \molsql. We then analyze how the gaps depend
% on the language of each component, on the difficulty level of the entity
% mappings, and on the data source, examine which method components narrow
% them, and close with an error analysis.

\subsection{Experimental Setup}
\label{sec:exp:setup}

\input{table/icde/tab7a_models}

\textbf{Datasets.} Unless noted otherwise, we evaluate on \molcube (480
instances $\times$ 8 cells, Section~\ref{sec:dataset:stats}), where the
languages of the question, the schema, and the values are crossed. Because
\molcube is source-balanced and enriched for language-changing mappings, its
magnitudes are diagnostic rather than deployment estimates. \molfull (4{,}223
instances $\times$ 4 cells) keeps the source distribution and is used for the
per-source accuracy and for the comparison with a joint switch of schema and
values.

\textbf{Models.} We select models that differ in access, size, and
architecture (Table~\ref{tab:models}):
\begin{itemize}
\item \emph{Open-weight models}: Qwen3.6-27B (Qwen-27B, dense),
Qwen3.6-35B-A3B (Qwen-35B, small MoE), DeepSeek-V4-Flash (DS-Flash,
284B/13B active), Qwen3.5-397B-A17B (Qwen-397B, large MoE), and
DeepSeek-V4-Pro (DS-Pro, 1.6T/49B active).
\item \emph{Proprietary models}: GPT-5.5, Claude Sonnet
4.6 (Claude-4.6), and Gemini 3.1 Pro (Gemini-3.1).
\end{itemize}
All eight models run the same few-shot prompt (FS), so that differences
among them reflect model capability alone. Qwen-35B and DS-Flash, two
open-weight models from different families with small active size, serve as
the backbones of all methods (the DB-grounded pipeline runs on Qwen-35B
only), which keeps the full method comparison affordable and reproducible.

\textbf{Methods.} We group the methods by the knowledge they can use to
bridge an entity mapping:
\begin{itemize}
\item \emph{Generation only}: zero-shot (ZS), five fixed English Spider
demonstrations (FS), zero-shot chain-of-thought (ZS-CoT), and retrieved
demonstrations with chain-of-thought (DFS-CoT).
\item \emph{Language alignment}: QTrans translates the question into the
language of the database and then generates, in the same call.
\item \emph{Value access}: DFS-CoT+VR adds to DFS-CoT only the cell values
that DeepEye-SQL's retriever~\cite{li2026deepeyesql} returns from the target
database.
\item \emph{DB-grounded pipeline}: DeepEye-L runs full DeepEye-SQL (schema
linking, value retrieval, three generators, revision, selection) with
sampling budget~1 per stage; DeepEye-L+DFS adds retrieved demonstrations.
\end{itemize}

\textbf{Metrics.} Execution accuracy (EX) is the primary metric. To locate
the source of an accuracy drop, we report the \emph{language gap} of each
component, $\Delta_Q$, $\Delta_S$, and $\Delta_V$ (Eq.~\eqref{eq:effect}): EX
with that component in English minus EX with it in Chinese, averaged over the
languages of the other two components. A positive gap means that the English
version is easier. 
In addition, two diagnostic measures are introduced where they are
used: the bridging cost (Section~\ref{sec:exp:difficulty}) and the literal
grounding accuracy (Section~\ref{sec:exp:components}).

% \textbf{Implementation details.} All calls use temperature~0 and a
% 4{,}096-token output budget. GPT-5.5 and Gemini-3.1 use low reasoning effort, because default reasoning exhausts the budget, and Claude-4.6 runs without extended thinking. 
% Strata use the labels of
% Section~\ref{sec:dataset:quality} and are reported only with at least 20
% instances.

\input{table/icde/tab7b_methods}

\begin{figure*}[t]
\centering
\includegraphics[width=\textwidth]{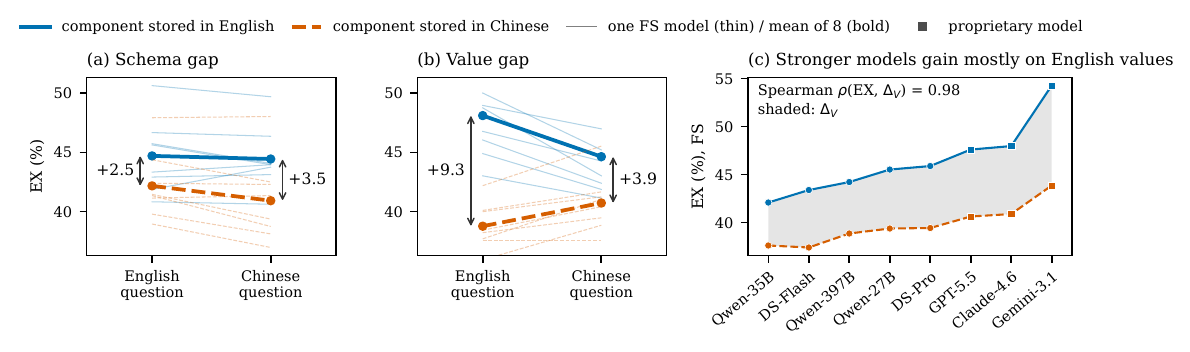}
\caption{The schema gap is the same under both question languages, whereas
the value gap depends on the question language (FS, \molcube). (a)(b) EX of
the $2{\times}2$ question-by-component cells, averaging over the third
component; arrows: gap between the two stored languages under each question
language. (c) EX on the $V_{\mathrm{en}}$ and $V_{\mathrm{zh}}$ cells for
each model, ordered by EX.}
\label{fig:decomposition}
\end{figure*}

\subsection{Main Results}
\label{sec:exp:main}

Tables~\ref{tab:models} and~\ref{tab:methods-results} report the results
of the models and of the methods.

\textbf{Stronger models widen the value gap.} Under FS, EX ranges from $39.8$ to $49.1\%$
(Table~\ref{tab:models}). The proprietary models lead the open-weight
models, but within the open-weight group size matters little: the dense
Qwen-27B outperforms both Qwen-35B and the far larger Qwen-397B. Every model
has a positive value gap, and the gap grows with model strength: across the
eight models, EX and $\Delta_V$ have Spearman $\rho=0.98$ ($0.85$ without
code-switched instances). Figure~\ref{fig:decomposition}(c) shows where the
gain goes: from the weakest to the strongest model, EX rises by $12$ points
on the $V_{\mathrm{en}}$ cells but only by $6$ on the $V_{\mathrm{zh}}$
cells, so the proprietary models combine the highest EX with the largest
$\Delta_V$. Scale thus helps mainly where the stored value is English, plausibly because the
model already knows that canonical form; it does not tell the model which
Chinese rendering a given database stores (Section~\ref{sec:exp:errors}).
The English demonstrations of FS are not the cause: ZS and FS have nearly
the same $\Delta_V$ (Table~\ref{tab:methods-results}).

\textbf{Database access matters more than model strength.} On the two backbones
(Table~\ref{tab:methods-results}), translating the question (QTrans) does
not raise EX over FS. Giving the generator target-database values lifts
DFS-CoT+VR to $51.2\%$ on DS-Flash, so a small-active-size open model with
value access already matches the best proprietary model under FS
(Gemini-3.1, $49.1\%$). The DB-grounded pipeline is best at $55.4\%$. At least $87\%$ of
the predictions of every run execute, so the errors lie in linking entities
rather than in producing runnable SQL, and even the best method fails on
almost half of the realizations.

\subsection{Impact of Component Language}
\label{sec:exp:language}

Switching the schema and the values separately shows that the two gaps have
different causes (Figure~\ref{fig:decomposition}(a)(b)).

\textbf{The schema gap does not depend on the question language.}
$\Delta_S$ is positive for every model under FS, but making the question
language match the schema language changes EX by only $-1.9$ to $+0.5$
points: the schema lines in Figure~\ref{fig:decomposition}(a) are parallel.
A Chinese question is even answered better over an English schema than over
a Chinese one, on all eight models. Chinese
identifiers are thus harder whatever the question language, consistent with
models having learned SQL mostly over English identifiers. Translating the
question cannot fix this; schema linking or identifier aliases target it
directly.

\textbf{The value gap depends on the direction of bridging.} The value
lines in Figure~\ref{fig:decomposition}(b) spread apart. Comparing the same
question over the two value languages, EX drops by $10.1$ to $23.3$ points
when an English mention must be matched to a Chinese stored value, but
changes by only $-1.6$ to $+4.5$ points in the reverse direction. $\Delta_V$ is therefore large
because the two directions are asymmetric, not because Chinese values are
harder in themselves. A plausible cause is that an English term often has
several acceptable Chinese renderings, of which the database stores one,
whereas a Chinese mention usually maps to a canonical English name.

\textbf{A joint database switch hides the difference.} On \molfull, where
schema and values switch together, DFS-CoT shows a joint database gap
$\Delta_{DB}\approx+9$ against $\Delta_Q\approx+2$--$3$ on both
backbones. This
single number folds together a schema gap and a value gap that call for
different remedies (Section~\ref{sec:problem:config}).

\begin{figure*}[t]
\centering
\includegraphics[width=\textwidth]{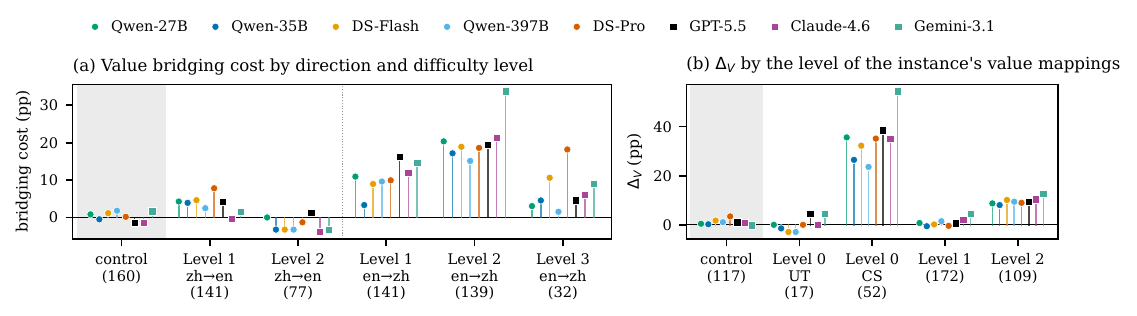}
\caption{Value-side cost of cross-lingual bridging on \molcube{} under FS,
with logical instance counts in parentheses (UT: same value in both
languages; CS: code-switched).}
\label{fig:bridging}
\end{figure*}

\input{table/icde/tab9_source}

\subsection{Impact of Mapping Difficulty}
\label{sec:exp:difficulty}

To attribute the value gap to the difficulty levels of
Section~\ref{sec:problem:level}, we measure a \emph{bridging cost}. For a
component $c\in\{S,V\}$, we take two realizations of instance $i$ that
differ only in the language $\ell_c$, such that a mapping of $c$ must be
bridged in $L'$ (mention and stored form in different languages) but no
mapping of $c$ needs bridging in $L$:
\begin{equation}
\label{eq:bridging}
B^{c}_{d,t}=\mathbb{E}_{(i,L,L')}\bigl[\mathrm{EX}_i(L)-\mathrm{EX}_i(L')\bigr],
\end{equation}
where $d$ is the direction of the bridged mapping (mention language $\to$
stored language) and $t$ its difficulty level.

\textbf{Without bridging, switching a language costs nothing.} When neither
realization needs bridging, the same difference serves as a control: had
migration translated an entity inconsistently or changed the task, it would
be nonzero. Where neither value cell needs bridging, it lies between
$-1.6$ and $+3.1$ points across all 20 runs (control in
Figure~\ref{fig:bridging}(a)), and $\Delta_V$ stays near zero on instances
without value mappings (Figure~\ref{fig:bridging}(b)). The costs below therefore come from
bridging rather than from migration artifacts.

\textbf{Convention-dependent mappings cost the most.} From English to
Chinese, Level-2 mappings cost $+15.1$ to $+33.7$ points, well above
Level-1 mappings (Figure~\ref{fig:bridging}(a)). Level-2 instances are not
simply harder queries: after direct standardization to the pooled
SQL-difficulty distribution, Level~2 still costs $+9.9$ to $+22.8$ and
exceeds Level~1 on all eight models. A Level-1 mapping
can be recovered from a dictionary gloss, whereas a Level-2 mapping requires
knowing which convention, such as an official name or a transliteration,
the data owner followed. English-to-Chinese Level-3 value mappings (32
instances) are too rare for a reliable estimate.

\textbf{Code-switched questions produce the largest gaps.} 52 Chinese
questions keep the English stored value verbatim (e.g., ``JetBlue
Airways''), as real users do. Their value mappings are at Level~0 in the
$V_{\mathrm{en}}$ cells but must be bridged in the $V_{\mathrm{zh}}$ cells
for both question languages. On these instances $\Delta_V$ reaches $+23.6$
to $+54.3$ (Figure~\ref{fig:bridging}(b)); without them the pooled
$\Delta_V$ drops to $+1.8$ to $+5.1$.

\subsection{Impact of Data Source}
\label{sec:exp:source}

Table~\ref{tab:source} breaks the results down by the five source
benchmarks.

\textbf{Accuracy varies more across sources than across methods.} On
\molcube, EX differs by about $50$ points between Spider, the easiest
source, and BULL, the hardest, and even the DB-grounded pipeline stays below
$42\%$ on EHRSQL and BULL. On \molfull, which keeps each source's own
instances, accuracy is clearly higher on BIRD, EHRSQL, and BULL but similar
on Spider and KaggleDBQA, so the per-source magnitudes on \molcube reflect
its enriched selection and should not be read as deployment estimates.

\textbf{Method components help on different sources.} Retrieved
demonstrations with chain-of-thought (DFS-CoT over FS) help most on the
domain-specific EHRSQL and BULL, where the fixed Spider demonstrations do not
show the domain's SQL patterns. Value retrieval (DFS-CoT+VR over DFS-CoT)
instead helps most on Spider and KaggleDBQA and not on BULL. Question
translation helps on Spider but lowers EX on KaggleDBQA.

\textbf{The schema gap is stable across sources; the value gap is not.}
Under FS, $\Delta_S$ is positive on every source for both backbones, whereas
$\Delta_V$ changes in size and sign. Spider has the largest $\Delta_V$,
mostly from its 30 code-switched instances (without them, $+8.3$ and
$+12.5$ instead of $+16.7$ and $+22.1$), and KaggleDBQA and BIRD lie in
between. BULL reverses ($\Delta_V$ negative on all eight models), also
without instances with a question--map inconsistency
(Section~\ref{sec:dataset:quality}); matching the question language to the
value language still helps (e.g., $+5.5$ points on DS-Flash), so the
reversal comes from Chinese stored values being easier overall, plausible
for a Chinese financial source. On EHRSQL, $\Delta_V$ is near zero because
two directions cancel, as surface opacity (Section~\ref{sec:problem:level})
predicts: English clinical abbreviations are opaque, while their Chinese
renderings spell out the meaning, and the bridging cost of its Level-1 value
mappings is indeed negative on seven of eight models. Only DeepEye-L+DFS
keeps $\Delta_V$ small on every source (Table~\ref{tab:source}), so the
DB-grounded pipeline also removes most of the source heterogeneity.

\begin{figure*}[t]
\centering
\includegraphics[width=\textwidth]{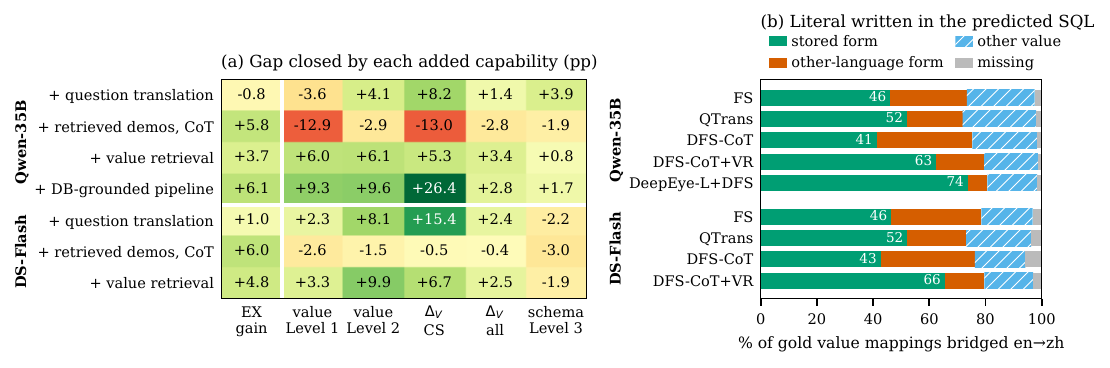}
\caption{Which method component closes which gap (\molcube). (a) Paired
difference between a method and the method it extends (QTrans$-$FS,
DFS-CoT$-$FS, DFS-CoT+VR$-$DFS-CoT, DeepEye-L+DFS$-$DFS-CoT+VR); Level
1/2/3: English-to-Chinese bridging cost; cost and $\Delta_V$ columns show
the reduction, so positive is better. (b) What
the predicted SQL wrote for value mappings bridged from English to Chinese;
numbers: LGA (\%, a lower bound).}
\label{fig:remedy}
\end{figure*}

\subsection{Impact of Method Components}
\label{sec:exp:components}

Section~\ref{sec:problem:level} predicts that translation suffices at
Level~1, model or domain knowledge at Level~2, and database lookup at
Level~3. We test this one component at a time by pairing each method with
the method it extends (Figure~\ref{fig:remedy}(a); Qwen-35B / DS-Flash). To
see whether a method writes the right stored value, we also measure the
\emph{literal grounding accuracy} (LGA): for each gold value mapping that
must be bridged, we check whether the predicted SQL writes the form stored in
that cell, the other language's form, another value, or no literal. LGA is a
lower bound, because an equivalent query that avoids the literal counts as
unmatched.

\textbf{Translation fixes code-switching but not the stored form.} QTrans
over FS shrinks $\Delta_V$ on code-switched instances and lowers the
English-to-Chinese Level-2 cost, but does not raise EX. Wrong-language
literals fall only from $27$--$32\%$ to $20$--$21\%$
(Figure~\ref{fig:remedy}(b)): translation turns the English mention into
Chinese, but it cannot tell which Chinese rendering the database stores.

\textbf{Demonstrations and reasoning raise EX but not the gap.} DFS-CoT over
FS adds about $6$ EX points on both backbones, yet $\Delta_V$ stays the same
or grows, wrong-language literals do not decrease, and on Qwen-35B the
English-to-Chinese Level-1 cost even grows. The added
component plausibly improves query structure, not literal grounding.

\textbf{Value retrieval narrows the gap.} DFS-CoT+VR over DFS-CoT raises EX
mostly on the $V_{\mathrm{zh}}$ cells, lowers the English-to-Chinese Level-2
cost, lifts LGA (Figure~\ref{fig:remedy}(b)), and halves wrong-language
literals (from $33$--$34\%$ to $14$--$17\%$).

\textbf{The DB-grounded pipeline removes most of the directional cost.}
DeepEye-L+DFS over DFS-CoT+VR (Qwen-35B) further raises EX and lowers the
English-to-Chinese Level-1 and Level-2 costs; wrong-language literals fall
to $7\%$, leaving $\Delta_V=+1.0$. Without a module-wise ablation, the gain
cannot be attributed to one component; demonstrations alone (DeepEye-L+DFS
over DeepEye-L) raise EX but leave $\Delta_V$ unchanged
(Table~\ref{tab:methods-results}).

\textbf{The prediction, revisited.} The Level-1 prediction is only partly
supported: translation acts mainly on code-switched mentions. The Level-2
prediction is not supported: reasoning, demonstrations, and stronger models
leave the Level-2 cost in place, whereas value retrieval reduces it, because
knowing how a term translates is not knowing how this database stores it.
The Level-3 prediction remains open: value-side Level-3 mappings are too
rare, and no method reliably lowers the schema-side English-to-Chinese
Level-3 cost.

\subsection{Error Analysis and Case Study}
\label{sec:exp:errors}

For value mappings that must be bridged from English to Chinese, the
literals written by the eight models under FS fall into three groups
(Figure~\ref{fig:remedy}(b)):
\begin{itemize}
\item \emph{Stored form} ($44$--$52\%$ of mappings): the mapping is bridged.
\item \emph{Other-language form} ($27$--$36\%$): the model keeps the English
value, e.g., \texttt{'JetBlue Airways'} against a Chinese database; this is
the error that value retrieval halves.
\item \emph{Other Chinese rendering} ($18$--$24\%$): the model translates the
mention but picks a form the database does not store, the Level-2 failure
that translation cannot fix.
\end{itemize}
In the reverse direction, GPT-5.5 and Claude-4.6 write the other language's
form far less often ($7\%$ and $9\%$, versus $17$--$21\%$ for the
open-weight models): model strength improves the Chinese-mention side, which
is why stronger models widen $\Delta_V$ (Section~\ref{sec:exp:main}).

Two cases illustrate the mechanisms.
\emph{Case 1 (Spider, code-switched brand name).} The Chinese question
\zh{航空公司}``JetBlue Airways''\zh{属于哪个国家？} keeps the brand name, and
the $V_{\mathrm{zh}}$ database stores \texttt{'\zh{捷蓝航空}'}. On Qwen-35B,
FS, DFS-CoT, QTrans, and DFS-CoT+VR all write \texttt{'JetBlue Airways'} and
fail, although each answers the same question over $V_{\mathrm{en}}$; only
DeepEye-L+DFS, which checks the database, writes \texttt{'\zh{捷蓝航空}'}.
\emph{Case 2 (BULL, established domain term).} The English question asks for
the ``Pharmaceutical Manufacturing Industry'', stored as
\texttt{'\zh{医药制造业}'}. On DS-Flash, DFS-CoT copies the English term,
while DFS-CoT+VR (by retrieval) and QTrans (by translation) both write the
stored form. A term with an established rendering can thus be bridged either
way, whereas a database-specific rendering needs the database.

%% file: table/icde/tab7a_models.tex
% Table VII-a (Sec. V-B). Generated by scripts/build_results.py from ci_results.json and literal_results.json.
\begin{table}[t]
\centering
\footnotesize
\setlength{\tabcolsep}{2.6pt}
\caption{Model comparison under FS on \molcube{}.}
\label{tab:models}
\begin{tabular}{@{}l rr r rrr@{}}
\toprule
\textbf{Model} & \textbf{Total} & \textbf{Active} & \textbf{EX} & $\Delta_Q$ & $\Delta_S$ & $\Delta_V$ \\
\midrule
\multicolumn{7}{@{}l}{\emph{Open-weight}} \\
\quad Qwen3.6-27B & 27B & dense & 42.4 & -0.4 & +2.4 & +6.1 \\
\quad Qwen3.6-35B-A3B & 35B & 3B & 39.8 & +0.9 & +1.8 & +4.5 \\
\quad DeepSeek-V4-Flash & 284B & 13B & 40.4 & +0.1 & +4.8 & +6.0 \\
\quad Qwen3.5-397B-A17B & 397B & 17B & 41.5 & +1.2 & +3.0 & +5.4 \\
\quad DeepSeek-V4-Pro & 1.6T & 49B & 42.7 & +1.9 & +4.5 & +6.5 \\
\multicolumn{7}{@{}l}{\emph{Proprietary}} \\
\quad GPT-5.5 & -- & -- & 44.1 & +1.8 & +1.4 & +7.0 \\
\quad Claude Sonnet 4.6 & -- & -- & 44.4 & +0.2 & +4.2 & +7.1 \\
\quad Gemini 3.1 Pro & -- & -- & \textbf{49.1} & +0.4 & +2.2 & +10.4 \\
\bottomrule
\end{tabular}
\end{table}

%% file: table/icde/tab7b_methods.tex
% Table VII-b (Sec. V-B). Generated by scripts/build_results.py from ci_results.json and literal_results.json.
\begin{table}[t]
\centering
\footnotesize
\setlength{\tabcolsep}{3.6pt}
\caption{Method comparison on \molcube{} with two backbone models. --: not
run because of the compute budget (Section~\ref{sec:exp:setup}).}
\label{tab:methods-results}
\begin{tabular}{@{}l rrrr rrrr@{}}
\toprule
 & \multicolumn{4}{c}{\textbf{Qwen-35B}} & \multicolumn{4}{c}{\textbf{DS-Flash}} \\
\cmidrule(lr){2-5}\cmidrule(lr){6-9}
\textbf{Method} & \textbf{EX} & $\Delta_Q$ & $\Delta_S$ & $\Delta_V$ & \textbf{EX} & $\Delta_Q$ & $\Delta_S$ & $\Delta_V$ \\
\midrule
\multicolumn{9}{@{}l}{\emph{Generation only}} \\
\quad ZS & 36.5 & +1.6 & +4.0 & +4.8 & 38.4 & +1.0 & +4.4 & +6.2 \\
\quad FS & 39.8 & +0.9 & +1.8 & +4.5 & 40.4 & +0.1 & +4.8 & +6.0 \\
\quad ZS-CoT & 40.8 & +1.6 & +0.6 & +4.7 & 38.6 & +2.5 & +2.4 & +6.0 \\
\quad DFS-CoT & 45.6 & +0.9 & +2.3 & +7.2 & 46.4 & +2.1 & +4.0 & +6.4 \\
\multicolumn{9}{@{}l}{\emph{Language alignment}} \\
\quad QTrans & 39.0 & +0.7 & +1.4 & +3.1 & 41.4 & +1.7 & +3.4 & +3.6 \\
\multicolumn{9}{@{}l}{\emph{Value access: target-database values}} \\
\quad DFS-CoT+VR & 49.3 & +2.3 & +1.7 & +3.8 & \textbf{51.2} & +3.6 & +5.0 & +3.9 \\
\multicolumn{9}{@{}l}{\emph{DB-grounded pipeline}} \\
\quad DeepEye-L & 49.9 & +0.5 & +0.6 & +1.0 & -- & -- & -- & -- \\
\quad DeepEye-L+DFS & \textbf{55.4} & +0.3 & +0.5 & +1.0 & -- & -- & -- & -- \\
\bottomrule
\end{tabular}
\end{table}

%% file: table/icde/tab9_source.tex
% Table IX (Sec. V-E). Generated by scripts/build_results.py from ci_results.json and ci_results_full.json.
\begin{table*}[t]
\centering
\footnotesize
\setlength{\tabcolsep}{3.6pt}
\caption{Results by data source.}
\label{tab:source}
\begin{tabular}{@{}l l r rrrrr rr rr@{}}
\toprule
 & & \textbf{Full EX} & \multicolumn{5}{c}{\textbf{\molcube{} EX}} & \multicolumn{2}{c}{\textbf{FS gaps}} &
 \multicolumn{2}{c}{\textbf{$\Delta_V$}} \\
\cmidrule(lr){3-3}\cmidrule(lr){4-8}\cmidrule(lr){9-10}\cmidrule(l){11-12}
\textbf{Source} & \textbf{Model} & DFS-CoT & FS & DFS-CoT & QTrans & +VR & DeepEye & $\Delta_S$ & $\Delta_V$ & +VR & DeepEye \\
\midrule
Spider & Qwen-35B & 70.0 & 71.6 & 66.3 & 74.5 & 75.4 & 85.7 & +1.0 & +16.7 & +13.8 & +4.2 \\
 & DS-Flash & 64.4 & 65.5 & 66.5 & 71.2 & 77.0 & -- & +6.0 & +22.1 & +9.6 & -- \\
\addlinespace[2pt]
KaggleDBQA & Qwen-35B & 49.5 & 47.1 & 51.0 & 41.7 & 57.6 & 62.0 & +1.6 & +3.6 & +1.0 & -1.0 \\
 & DS-Flash & 46.7 & 49.1 & 48.7 & 45.8 & 55.1 & -- & +4.4 & +6.0 & +1.3 & -- \\
\addlinespace[2pt]
BIRD & Qwen-35B & 42.6 & 32.3 & 36.3 & 32.0 & 39.6 & 52.6 & +3.1 & +5.7 & +4.7 & +3.1 \\
 & DS-Flash & 41.9 & 31.2 & 33.2 & 32.6 & 39.1 & -- & +4.7 & +3.6 & +2.6 & -- \\
\addlinespace[2pt]
EHRSQL & Qwen-35B & 51.2 & 28.6 & 44.9 & 27.9 & 44.9 & 41.5 & +2.1 & +0.5 & +1.8 & +0.8 \\
 & DS-Flash & 54.3 & 34.6 & 49.9 & 34.6 & 53.6 & -- & +5.2 & +2.1 & +3.6 & -- \\
\addlinespace[2pt]
BULL & Qwen-35B & 41.0 & 19.5 & 29.4 & 19.1 & 29.0 & 35.4 & +1.0 & -4.2 & -2.3 & -2.1 \\
 & DS-Flash & 43.0 & 21.5 & 33.7 & 22.9 & 31.2 & -- & +3.9 & -3.9 & +2.1 & -- \\
\midrule
Pooled & Qwen-35B & 52.3 & 39.8 & 45.6 & 39.0 & 49.3 & 55.4 & +1.8 & +4.5 & +3.8 & +1.0 \\
 & DS-Flash & 52.3 & 40.4 & 46.4 & 41.4 & 51.2 & -- & +4.8 & +6.0 & +3.9 & -- \\
\bottomrule
\end{tabular}
\par\vspace{3pt}
\end{table*}

% \parbox{\textwidth}{\scriptsize +VR: DFS-CoT+VR; DeepEye: DeepEye-L+DFS, run on Qwen-35B only. \molcube{} is balanced and challenge-enriched; magnitudes are diagnostic, not deployment estimates.}

%% file: src/related.tex
\section{Related Work}
\label{sec:related}

\paragraph{Text-to-SQL Benchmarks}
Spider established cross-domain Text-to-SQL evaluation, and BIRD added
large, value-rich databases~\cite{yu2019SpiderLargeScaleHumanLabeled,li2023CanLLMAlready};
KaggleDBQA, BULL, and EHRSQL expose real-world and domain-specific schemas,
terminology, and
values~\cite{lee2021KaggleDBQA,zhang2024FinSQLModelAgnosticLLMsbased,lee2026EHRSQLPracticalTexttoSQL}.
These benchmarks are monolingual. CSpider, MultiSpider, and DuSQL broaden
the question language but leave the database
unchanged~\cite{min2019PilotStudyChineseCspider,dou2023MultiSpiderBenchmarkingMultilingual,wang2020DuSQLLargeScalePragmatic}.
Coordinated localization goes further: PAUQ adapts questions, SQL, and
database content to Russian~\cite{bakshandaeva2022pauq}, MultiSpider~2.0
localizes schema and values with consistency
checks~\cite{pham2025MultilingualTexttoSQLBenchmarking}, and BIRDTurk~\cite{aktas2026BIRDTurk}
translates the schema while keeping values unchanged. Each, however, releases
one or a few coupled configurations, and newly constructed workloads such as
TACO~\cite{deng2026TACOBenchmarkOpenDomain} add a semantic-validation burden
of their own. \molsql instead migrates established workloads and switches
$Q$, $S$, and $V$ separately on the same logical instance to separate the
component effects, and it grades each entity mapping by difficulty.

\paragraph{Data Migration and Verified SQL Transformation}
Entity matching determines whether two records refer to the same real-world
entity~\cite{li2020DeepEntityMatching}, but identifying correspondences does
not transform the schema, values, constraints, gold SQL, and question of a
Text-to-SQL instance together. Schema evolution and database migration treat
a change to a database as a coupled update of schema, state, and dependent
artifacts~\cite{curino2008PRISM,hammer1975SemanticIntegrity,lee2025SemanticIntegrityConstraints},
and verified SQL transformation accepts a rewritten query only with
execution or equivalence evidence~\cite{zhou2025CrackingSQL,wang2024QED}.
\molmigrate applies these principles to linguistic rather than structural or
dialect transformation: one entity map governs every occurrence of an entity,
SQL-bearing constraints and queries are rewritten under resolved scope, and
every migrated query is accepted only after an execution check.

%% file: src/conclusion.tex
\section{Conclusion and Limitations}
\label{sec:conclusion}
\label{sec:exp:lessons}

We presented \molsql, a benchmark for Mixture-of-Language Text-to-SQL that
fixes each task and varies the language of the question, the schema, and the
stored values, together with \molmigrate, which builds aligned realizations
from five established workloads through one entity map and execution checks.
Experiments on its two splits show that the schema and value languages have
distinct effects that stem from cross-lingual entity mapping. Neither
question translation nor additional reasoning narrows the value gap, while
language-aware context grounding in the target database substantially
reduces it; the schema gap instead calls for schema linking or identifier
aliases.

\molsql covers only English--Chinese and SQLite, checks equivalence on a
single database state, and builds on public databases that models may have
seen during pretraining; \molcube is a diagnostic sample rather than a
deployment estimate. Extending the $Q/S/V$ design to more languages and DBMSs
would test how far these findings generalize.